\documentclass[letterpaper,twocolumn,pra,aps,superscriptaddress,floatfix,amsmath,amssymb]{revtex4}
\usepackage{graphicx}
\usepackage{multirow}
\usepackage{natbib}
\usepackage{array}
\usepackage{multirow,amssymb,amsbsy,amsmath,epstopdf}
\usepackage{color}

\usepackage[colorlinks,citecolor=blue]{hyperref}

\begin{document}

\title{Coherent control of defect spins in silicon carbide above 550 K}

\author{Fei-Fei Yan}
\affiliation{CAS Key Laboratory of Quantum Information, University of Science and Technology of China, Hefei 230026, People's Republic of China}
\affiliation{Synergetic Innovation Center of Quantum Information and Quantum Physics, University of Science and Technology of China, Hefei 230026, People's Republic of China}

\author{Jun-Feng Wang}
\affiliation{CAS Key Laboratory of Quantum Information, University of Science and Technology of China, Hefei 230026, People's Republic of China}
\affiliation{Synergetic Innovation Center of Quantum Information and Quantum Physics, University of Science and Technology of China, Hefei 230026, People's Republic of China}

\author{Qiang Li}
\affiliation{CAS Key Laboratory of Quantum Information, University of Science and Technology of China, Hefei 230026, People's Republic of China}
\affiliation{Synergetic Innovation Center of Quantum Information and Quantum Physics, University of Science and Technology of China, Hefei 230026, People's Republic of China}

\author{Ze-Di Cheng}
\affiliation{CAS Key Laboratory of Quantum Information, University of Science and Technology of China, Hefei 230026, People's Republic of China}
\affiliation{Synergetic Innovation Center of Quantum Information and Quantum Physics, University of Science and Technology of China, Hefei 230026, People's Republic of China}

\author{Jin-Ming Cui}
\affiliation{CAS Key Laboratory of Quantum Information, University of Science and Technology of China, Hefei 230026, People's Republic of China}
\affiliation{Synergetic Innovation Center of Quantum Information and Quantum Physics, University of Science and Technology of China, Hefei 230026, People's Republic of China}

\author{Wen-Zheng Liu}
\affiliation{CAS Key Laboratory of Quantum Information, University of Science and Technology of China, Hefei 230026, People's Republic of China}

\author{Jin-Shi Xu}\email{jsxu@ustc.edu.cn}
\affiliation{CAS Key Laboratory of Quantum Information, University of Science and Technology of China, Hefei 230026, People's Republic of China}
\affiliation{Synergetic Innovation Center of Quantum Information and Quantum Physics, University of Science and Technology of China, Hefei 230026, People's Republic of China}

\author{Chuan-Feng Li}\email{cfli@ustc.edu.cn}
\affiliation{CAS Key Laboratory of Quantum Information, University of Science and Technology of China, Hefei 230026, People's Republic of China}
\affiliation{Synergetic Innovation Center of Quantum Information and Quantum Physics, University of Science and Technology of China, Hefei 230026, People's Republic of China}

\author{Guang-Can Guo}
\affiliation{CAS Key Laboratory of Quantum Information, University of Science and Technology of China, Hefei 230026, People's Republic of China}
\affiliation{Synergetic Innovation Center of Quantum Information and Quantum Physics, University of Science and Technology of China, Hefei 230026, People's Republic of China}
\date{\today}

\begin{abstract}
Great efforts have been made to the investigation of defects in silicon carbide for their attractive optical and spin properties. However, most of the researches are implemented at low and room temperature. Little is known about the spin coherent property at high temperature. Here, we experimentally demonstrate coherent control of divacancy defect spins in silicon carbide above 550 K. The spin properties of defects ranging from room temperature to 600 K are investigated, in which the zero-field-splitting is found to have a polynomial temperature dependence and the spin coherence time decreases as the temperature increases. Moreover, as an example of application, we demonstrate a thermal sensing using the Ramsey method at about 450 K. Our experimental results would be useful for the investigation of high temperature properties of defect spins and silicon carbide-based broad-temperature range quantum sensing.

\end{abstract}
\maketitle



\section{Introduction}

Most recently, defects in silicon carbide (SiC) have attracted many attentions for their appealing optical and spin properties \citep{koehl2011room,falk2013polytype,christle2015isolated,fuchs2015engineering,
widmann2015coherent,castelletto2014silicon,lienhard2016bright,wang2017efficient,
christle2017isolated,klimov2015quantum,seo2016quantum,falk2014electrically,
klimov2014electrically,zhou2017self}. Since SiC is a widely used semiconductor material in electronic devices with outstanding features such as various mature fabrication technology and large-scale high quality single crystal growth, defect-based control and manipulation in SiC would lead to practical wafer-scale quantum information processing \citep{koehl2011room,falk2013polytype,christle2015isolated,fuchs2015engineering,widmann2015coherent}. Besides some types of high bright single photon sources \citep{castelletto2014silicon,lienhard2016bright}, there are also two types of spin-qubit in SiC: the silicon vacancy defect \citep{christle2015isolated,fuchs2015engineering,wang2017efficient} and divacancy defect consisting of a carbon vacancy adjacent to a silicon vacancy \citep{koehl2011room,falk2013polytype,christle2015isolated,christle2017isolated}. Similar to nitrogen vacancy (NV) centers in diamond \citep{maze2008nanoscale}, the defect spins in SiC can be polarized and readout by laser and controlled by microwave. Due to the outstanding properties of infrared (IR) fluorescence wavelength and long coherence time ($ \sim ms$), the divacancy defects with spin $S=1$ in SiC have enabled applications in quantum metrologies, including magnetic \citep{niethammer2016vector,simin2016all,simin2015high}, electric \citep{wolfowicz2018electrometry,falk2014electrically} and temperature sensing \citep{zhou2017self}, etc. Unlike other thermometers such as fluorescence-based organic dyes and Raman spectroscopy with the drawbacks of local environmental disturbance and low-sensitivity \citep{brites2012thermometry}, the SiC-based quantum sensor have advantages of photostable fluorescence, chemical inertness, non-toxic, and high sensitivity \citep{christle2015isolated,christle2017isolated,seo2016quantum,zhou2017self}. Moreover, the IR range of fluorescence may be more suitable for biological systems than that of NV centers in diamond \citep{zhou2017self,kucsko2013nanometre}.

However, most of the previous experiments are operated at low and room temperature. Little is known about the spin coherent property at high temperature. Understanding such kind of property is important for manipulating defect spins in SiC at high-temperature, which would be useful for making a broad-temperature range high sensitivity quantum sensor.

In this work, we study the coherent property of divacancy defect spins in 4H-SiC with the temperature ranging from 300 to 600 K. In particular, we focus on the PL5 divacancy defects with a basal $C_{1h}$ symmetry. By measuring the optically detected magnetic resonance (ODMR) spectrum, we find that the zero-field-splitting has a polynomial temperature dependence and the contrast of the ODMR signal decreases with the increase of temperature which almost vanishes at about 600 K. The spin coherent time is also found to decrease as the temperature increases. The infrared high temperature thermometry based on Ramsey methods is also demonstrated. The results pave the way for the SiC-based broad temperature range quantum sensing such as magnetic and temperature sensing.

\section{Experimental setup and results}
\begin{figure}[!htb]
\begin{center}
\includegraphics[width=0.95\columnwidth]{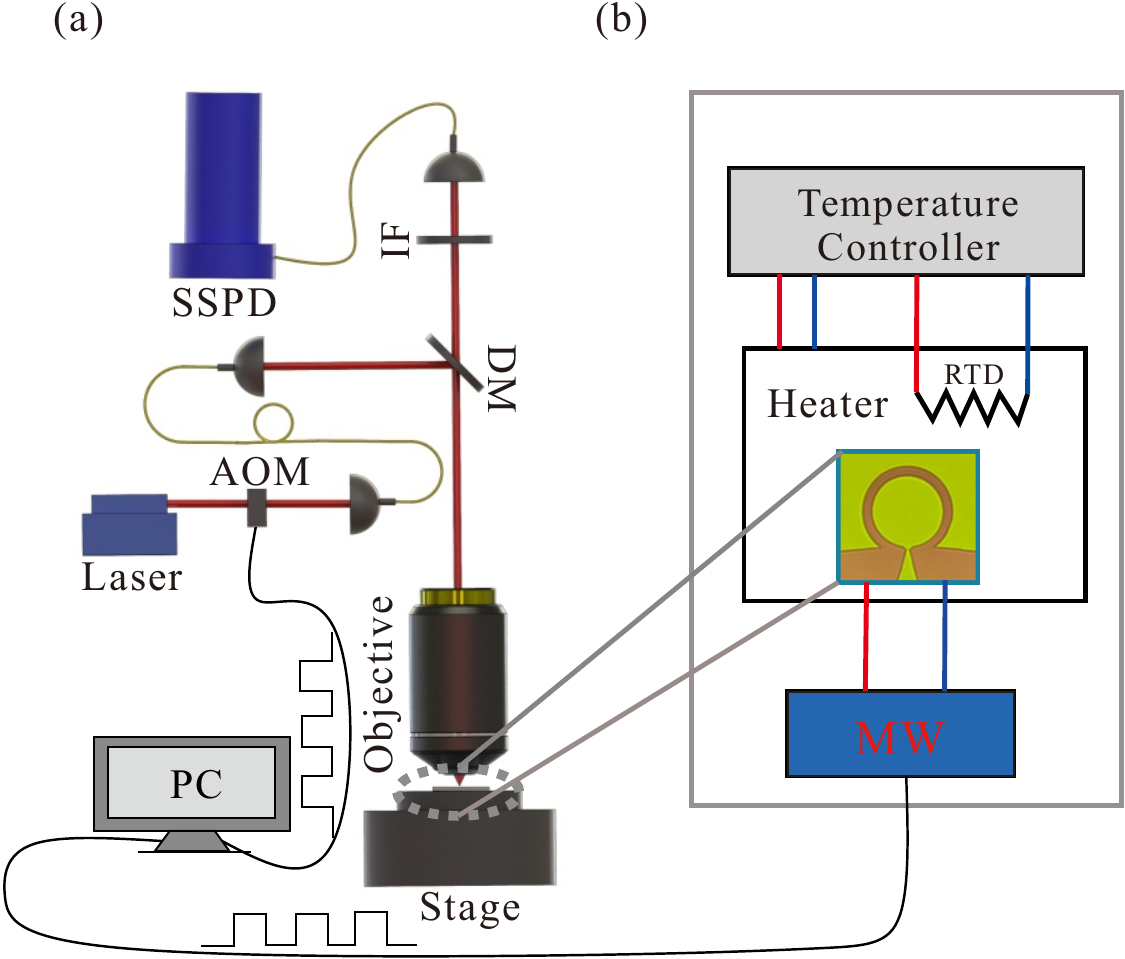}
\caption{Experimental setup. (a) The home-built confocal microscope system. A laser with the wavelength of 920 nm is modulated by an acousto-optic modulator (AOM). After reflected by a dichroic mirror (DM), laser pulses are focused by an objective to pump the sample. The fluorescence is collected by the same objective and filtered by an interference filter (IF), which is finally detected by a superconducting single photon detector (SSPD). (b) The sample is fixed on a metal ceramic heater and the temperature is detected by a resistive temperature detector (RTD), which are connected to a temperature controller. A ring shape antenna is mounted on the sample to implement microwave (MW) pluses. The electrical pulse sequences on the AOM and MW system are controlled by a computer (PC). }

\label{setup}
\end{center}
\end{figure}

In this work, we use a home-built confocal microscopy system and a microwave system to excite and control the defect spins in a commercial high-purity semi-insulating 4H-SiC sample purchased from Cree. As shown in Fig.~\ref{setup}(a), an IR laser with the center wavelength at 920 nm modulated by an acousto-optic modulator (AOM) is used to excite the defects. After reflected by a dichroic mirror (DM) (980 nm long-pass), the laser is focused on the SiC sample by a near-infrared objective (NA = 0.7) with a long work distance. The fluorescence is collected by the same objective and detected by a superconducting single photon detector (SSPD, Scontel) after a 1064 nm long-pass interference filter (IF). A ring-shaped microwave antenna is fabricated on the sample for transmitting microwave to manipulate defect spins (Fig.~\ref{setup}(b)). The electrical pulse sequences for manipulating the laser and microwave are generated by a pulse generator (PBESR-PRO-500, Spincore) in a computer (PC). In order to control the sample's temperature, the sample is mounted on a metal ceramic heater (HT24S, Thorlabs) with a resistive temperature detector (RTD) (TH100PT, Thorlabs) fixed on it. The heater and RTD are controlled by a temperature controller to achieve stability within $\pm 100$ mK up to 600 K in the atmospheric environment.


\begin{figure}[!htb]
\begin{center}
\includegraphics[width=0.95\columnwidth]{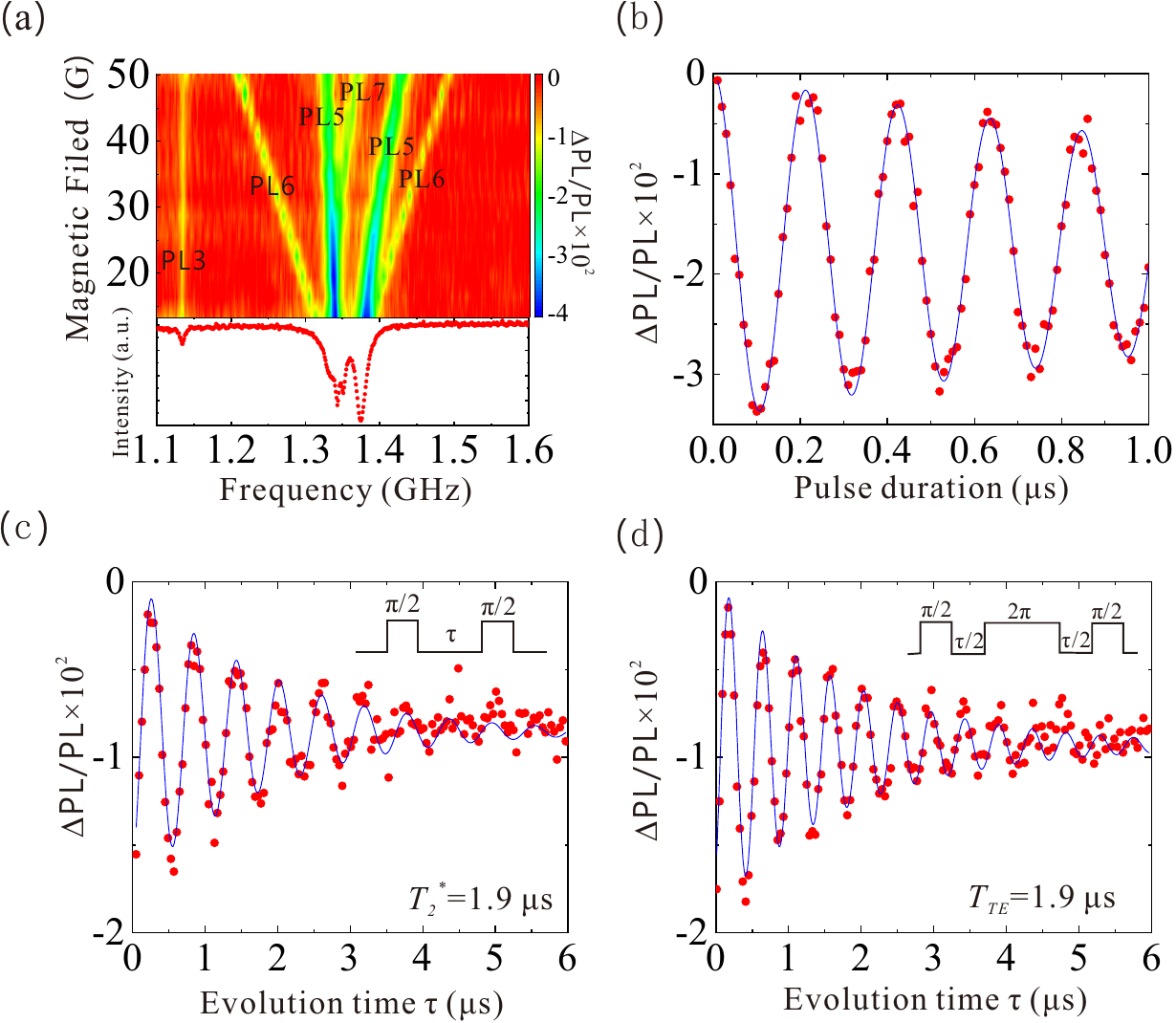}
\caption{Coherent controls of spins at room temperature. (a) The ODMR spectrum of divacancy defect spins in 4H-SiC with the increase of magnetic field (upper panel) and the zero magnetic field (down panel). The labeled PL3, 5-7 are four types of divacancies. (b) Rabi oscillation of PL5. The blue line represents the theoretical fit with an exponentially decaying sinuous function. (c) Ramsey fringe of PL5. The decay time shows that the dephasing time $T_{2}^{*}$ is about 1.9 $\mu s$. (d) Thermal echo oscillation of PL5. The thermal echo dephasing time $T_{TE}$ is also about 1.9 $\mu s$. The blue lines and insets in (c) and (d) are the fits using Eq.~\ref{fit} and the microwave pulse sequences, respectively.}

\label{room-temperature}
\end{center}
\end{figure}

To identify defects in the 4H-SiC sample, we scan the ODMR spectrum as a function of the applied magnetic field along the c axis of the sample, which is shown in the upper panel in Fig. \ref{room-temperature}(a) and agrees with the previous results \citep{falk2013polytype}. Four distinct divacancy defects are detectable at room temperature, which are labeled as PL3, 5, 7 with basal symmetry and PL6 with c axis symmetry. The two transitions of PL6 defects are measured to split at the slope of 2.8 MHz/G with the c-axis magnetic field \citep{koehl2011room,falk2013polytype}. The down panel in Fig. \ref{room-temperature}(a) shows ODMR spectrum at zero magnetic field. The zero-field-splitting (ZFS) parameter ($D$) of the right branch of PL5 is measured to be 1374.5 MHz, which further confirms the type of divacancy defects. Since it is easily distinguishable and has a higher contrast and long Ramsey coherence time, we would focus on the right branch of PL5 defects to investigate spin dynamics at high temperature with zero magnetic field.



Coherent controls of spins at room temperature including the implementation of Rabi oscillation, Ramsey and thermal echo \citep{kucsko2013nanometre,zhou2017self} measurements are shown in Fig. \ref{room-temperature}(b), (c) and (d), respectively. In the experiment, the length of the initialized and read-out laser pulses are both set to be 3 $\mu$s \citep{koehl2011room,falk2013polytype,zhou2017self}. The Rabi oscillation is measured as a function of the microwave pulse duration, as shown in Fig.~\ref{room-temperature}(b). The microwave pulse sequences for Ramsey and thermal echo oscillations are shown in the insets of Fig. \ref{room-temperature}(c) and (d), respectively. Blue lines are the theoretical fits. An exponentially decaying sinuous function is used to fit the Rabi oscillation. While the Ramsey and thermal echo oscillations are fitted with the function
\begin{equation}\label{fit}
I_{PL}=a\exp{(-(\frac{t}{T_{d}})^{n})} \cos{(2\pi ft+\varphi)}+b,
\end{equation}
\noindent where $a$, $n$, $b$ and $\varphi$ are free parameters. $T_{d}$ represents the dephasing time \citep{koehl2011room,falk2013polytype,zhou2017self}. The oscillating frequency $f$ corresponds to the detuning frequency from the ZFS parameter, which is set to be 1.7 MHz. The Ramsey dephasing time ($T_{2}^{*}$) and thermal echo dephasing time ($T_{TE}$) are both obtained to be about 1.9 $\mu s$, which are consistent with previous results \citep{koehl2011room,zhou2017self}. It is due to the fact that the PL5 has a large transverse strain E (about 13.5 MHz) which has a self-protected effect for the local slowly varying magnetic field in the Ramsey measurement, the coherence time is not expected to increase with thermal echo in our experiment~\cite{zhou2017self}. The coherence time ($T_{2}$) is measured to be about 48.9 $\mu s$ in our experiment.

\begin{figure}[!htb]
\begin{center}
\includegraphics[width=0.95\columnwidth]{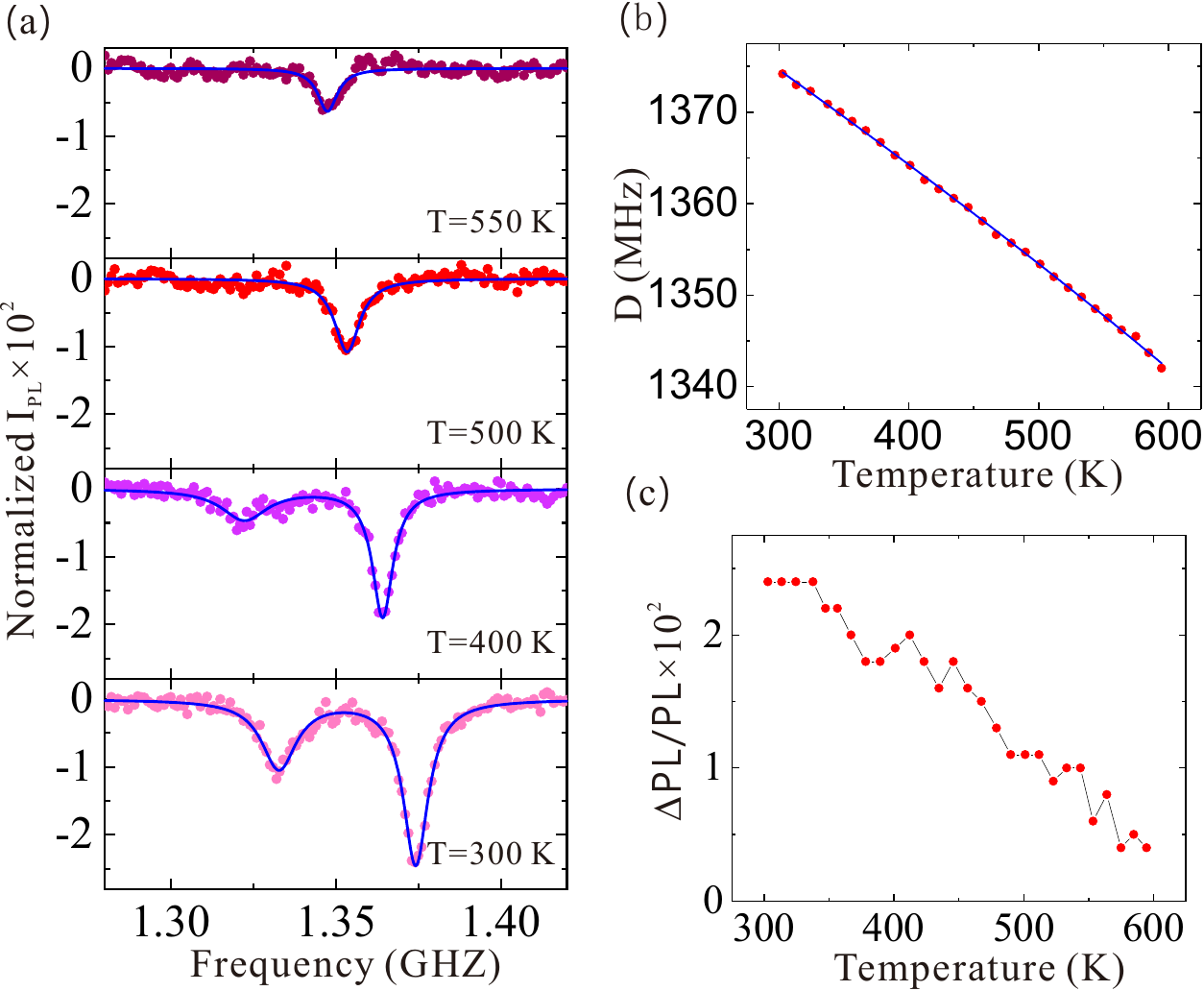}
\caption{Temperature dependence of ODMR signals for PL5. (a) The ODMR spectrum at four representative temperatures. The blue lines are the theoretical Lorentzian fits. (b) The ZFS parameter $D$ as a function of the temperature increasing from 300 to 600 K. The blue line is the theoretical fit with a third-order polynomial. The error bar of every point deduced from the ODMR fitting error is smaller than the corresponding symbol size. (c) The corresponding ODMR contrasts of the right branch of PL5 as a function of the temperature increasing from 300 to 600 K.}

\label{ODMR VS Tem}
\end{center}
\end{figure}

We then detect the ODMR spectrum at temperatures ranging from 300 to 600 K. Fig. \ref{ODMR VS Tem}(a) shows four representative ODMR signals at 300, 400, 500 and 550 K, respectively, which are fitted with Lorentzian functions (blue lines). Both the ZFS parameter $D$ and ODMR contrast decrease as the temperature increases. The details of the dependence of $D$ on temperature is shown in Fig. \ref{ODMR VS Tem}(b). The change of $D$ is about -32 MHz from 300 to 600 K. We use a three-order polynomial function to fit the data, which reads as $D(T)=a_{0}+a_{1}(T-300)+a_{2}(T-300)^{2}+a_{3}(T-300)^{3}$ with parameters $a_{0}$ = $1374.5\pm0.1$ MHz, $a_{1}$ = $-0.096\pm0.003$ MHz/K, $a_{2}$ = $(-8.1\pm3.3)\times 10^{-5}$ MHz/K$^{2}$ and $a_{3}$ = $(1.4\pm0.9)\times 10^{-7}$ MHz/K$^{3}$, respectively \citep{toyli2012measurement,plakhotnik2014all,li2017temperature}. Compared with the case of NV centers \citep{toyli2012measurement}, we obtain a better linear relationship, which would pave the way for constructing the ODMR signal based broad temperature thermal sensor. The dependence between the ODMR contrast and temperature is shown in Fig. \ref{ODMR VS Tem}(c). The decrease of the ODMR contrast might also be caused by the thermally activated nonradiative processes which would diminish the fluorescence-based spin readout~\citep{toyli2012measurement}.



\begin{figure}[!htb]
\begin{center}
\includegraphics[width=0.95\columnwidth]{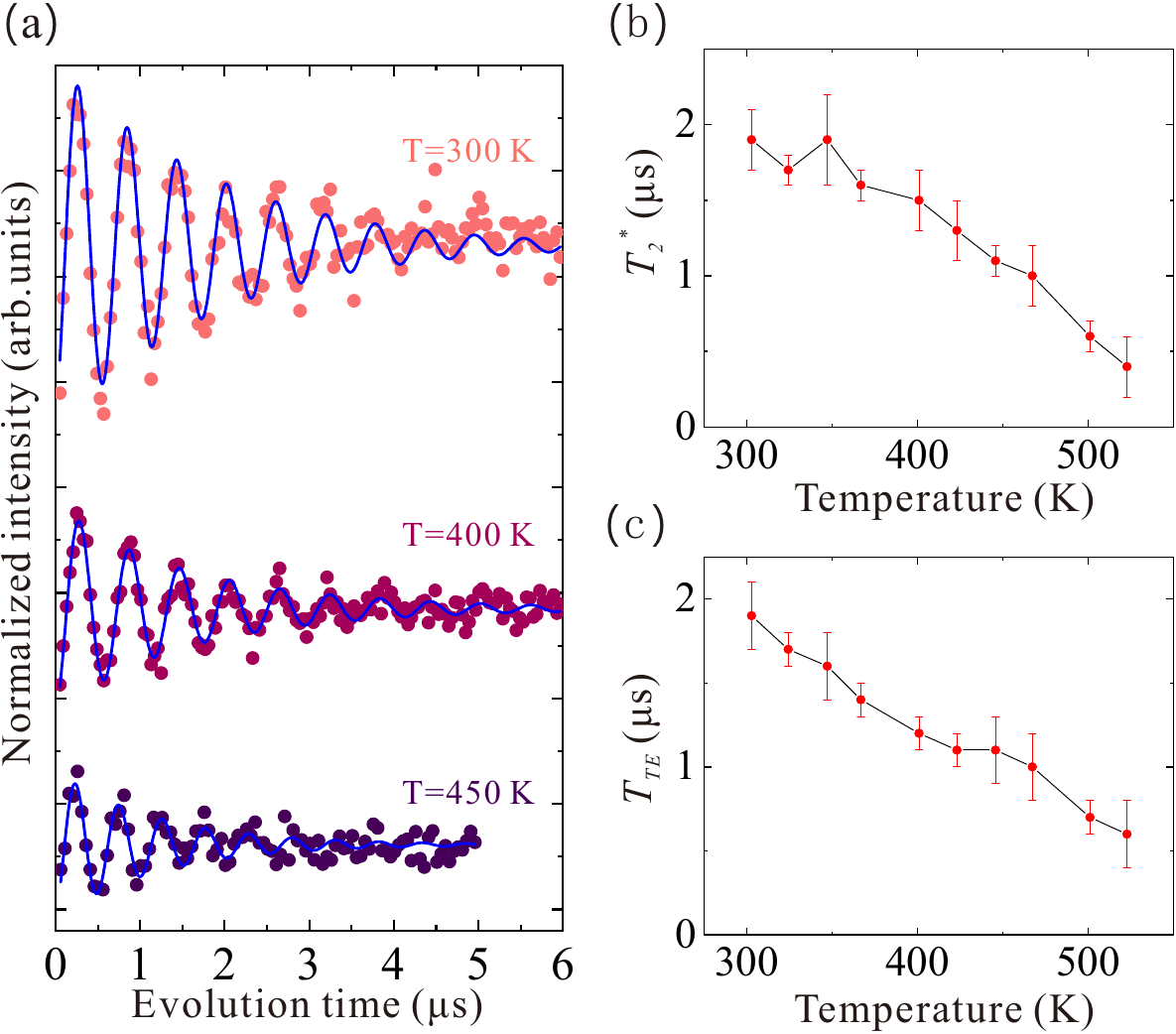}
\caption{Temperature dependence of coherence time measurement. (a) The Ramsey oscillations at three different temperatures. (b) The dephasing time $T_{2}^{*}$ versus temperature. (c) The thermal echo dephasing time $T_{TE}$ versus temperature. }
\label{dephasing time VS Tem}
\end{center}
\end{figure}

Ramsey and thermal echo oscillations are widely used in the quantum sensing \citep{taylor2008high,neumann2013high,schirhagl2014nitrogen,zhou2017self,wang2015high,
toyli2013fluorescence,kucsko2013nanometre}. We further investigate the temperature dependence of Ramsey and thermal echo oscillations. Fig. \ref{dephasing time VS Tem}(a) shows three Ramsey measurements at different temperature and Fig. \ref{dephasing time VS Tem}(b) shows the dephasing time $T_{2}^{*}$ varying with temperature ranging from 300 to 520 K. Similar with the temperature dependence of $D$, both the contrast of Ramsey oscillation and $T_{2}^{*}$ decrease as the temperature increases. In particular, $T_{2}^{*}$ reduces to only about 0.4 $\mu s$ at 520 K. Unlike the behavior of $T_{2}^{*}$ of NV spins in diamond which maintains almost the same from 300 to 620 K \citep{toyli2012measurement}, the $T_{2}^{*}$ of PL5 divacancy defects in 4H-SiC first decreases slowly when the temperature is below 400 K, then decreases quickly when the temperature is increased to about 520 K. Further investigations are needed in order to exactly understand the origin of the difference. Moreover, the coherence time of thermal echo oscillation $T_{TE}$ varying as a function of temperature is shown in Fig. \ref{dephasing time VS Tem}(c). The $T_{TE}$ decreases as the temperature increases which is similar with the results of $T_{2}^{*}$. $T_{TE}$ decreases to only about 0.6 $\mu s$ at 520 K.



\begin{figure}[!htb]
\begin{center}
\includegraphics[width=0.95\columnwidth]{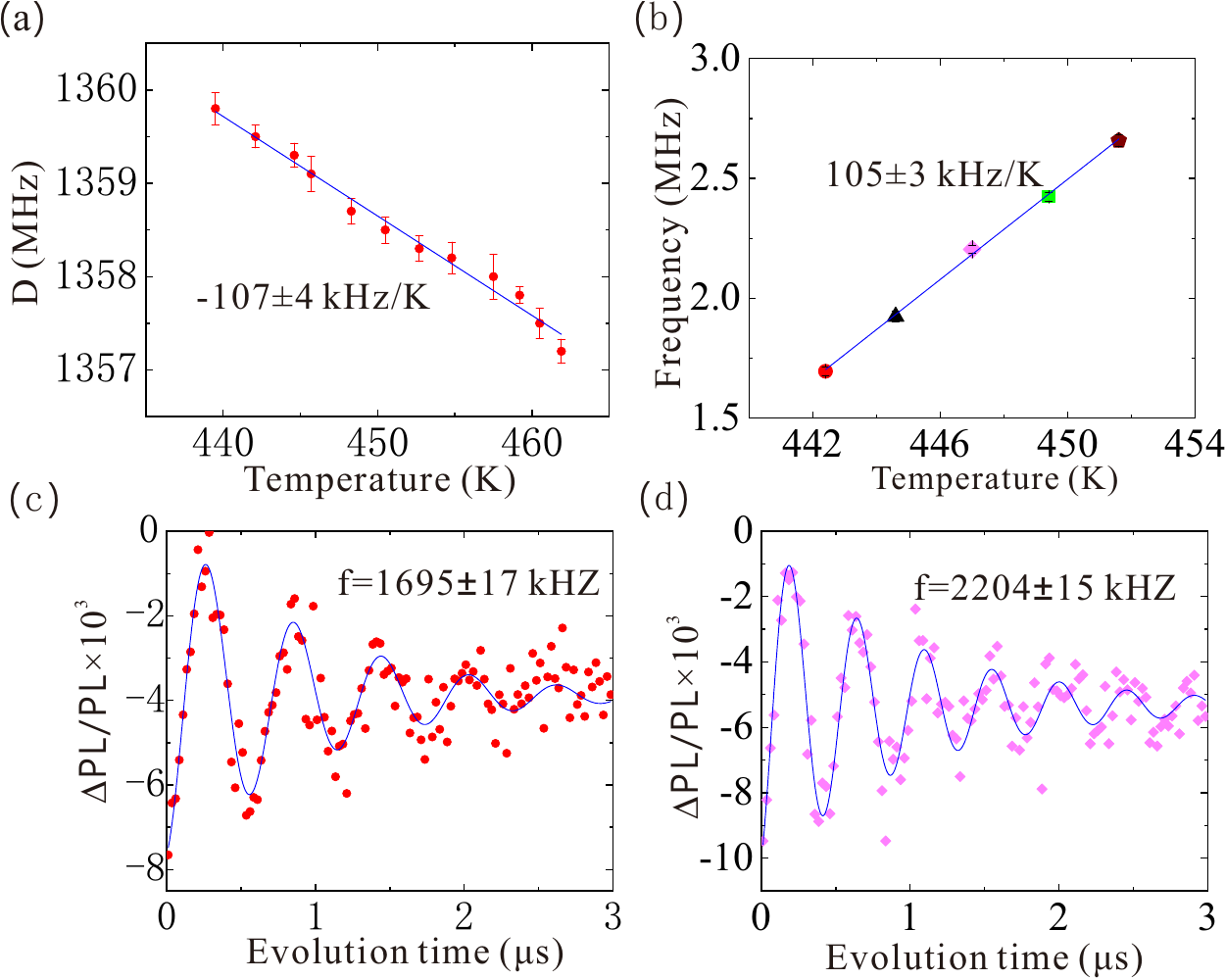}
\caption{Thermal sensing properties at about 450 K. (a) The ZFS parameter $D$, inferred from the Lorentzian fit to the ODMR data, at about 450 K. The blue line shows the linear fitting. (b) Ramsey oscillation frequency as a function of temperature. The blue line is a linear fitting. The slope of Ramsey oscillation frequency is consistent with the slope of $D$ as a function of temperature. (c) and (d) Ramsey oscillations at two different temperatures show different oscillation frequencies. Blue lines are the fittings using Eq. \ref{fit}. The corresponding oscillation frequencies are shown in (b) in red circle and pink rhombus, respectively. }

\label{Thermometer}
\end{center}
\end{figure}

The PL5 spin dynamics in 4H-SiC can be used for high-temperature quantum sensing \citep{toyli2012measurement,toyli2013fluorescence,anisimov2016optical}. We investigate a thermal sensing at about 450 K using the Ramsey method. As shown in Fig. \ref{Thermometer}(a), the ZFS parameter $D$ deduced from the ODMR spectrums ranging from 440 to 460 K, are shown to linearly decrease with the increase of temperature with a slope of $-107\pm4$ kHz/K. We then measure the frequencies of Ramsey oscillations at five different temperatures, which is shown in Fig. \ref{Thermometer}(b). The MW frequency is kept constant for all further measurements after it is set to 1.7 MHz detuning at 442.4 K. There is a linear relation between the oscillation frequency and temperature. The slope is measured to be $105\pm3$ kHz/K, which is consistent with that in Fig. \ref{Thermometer}(a). Two Ramsey oscillations at 442.4 K and 447 K are shown in Fig. \ref{Thermometer}(c) and (d) with the oscillation frequencies being $f=1695\pm 17$ kHz and $f=2204\pm 15$ kHz, respectively. The dephasing time $T_{2}^{*}$ is deduced to be about 1.1 $\mu s$. Using the standard error derived from the fitting, we estimate that the precision of measured temperature is about 160 mK. The sensitivity is a significant physical quantity to measure the quality of a thermometer. We can get the sensitivity using the equation
\begin{equation}\label{sensitivity}
\eta=\sqrt{\frac{2(p_{0}+p_{1})}{(p_{0}-p_{1})^{2}}} \frac{1}{2\pi \frac{dD}{dT} \exp{(-(\frac{t}{T_{d}})^{n})} \sqrt{t}},
\end{equation}
with $p_{0}$ and $p_{1}$ being the photon counts per measurement shot for the bright and dark spin states, respectively. The sensitivity $\eta$ value corresponds to the maximum value of $\exp{(-(\frac{t}{T_{d}})^{n})} \sqrt{t}$ \citep{zhou2017self,wang2015high}. In the experiment, the sensitivity is about 274 mK/Hz$^{1/2}$ at room temperature and about 880 mK/Hz$^{1/2}$ at 450 K, respectively. The sensitivity should be better with a higher collection efficiency. In our experiment, the counts is about 570 kcps at 4 mW laser pump at room temperature and is stable during the measurement. The reduced sensitivity at higher temperature is due to the decrease of photon counts, contrast of ODMR signal and $T_{2}^{*}$ as the temperature increases.



\section{Conclusions}
In summary, we have investigated the ODMR spectrum and Ramsey oscillation of the PL5 divacancy defect spins in 4H-SiC from 300 to 600 K. The results show that coherent controls on defect spins in 4H-SiC can be achieved even at high temperature up to 600 K. As an application, a high-temperature thermal sensing at 450 K is demonstrated with the PL5 divacancy defect spins, which are shown to be useful for constructing a robust broad temperature range thermometer with a high sensitivity. Temperature sensitivity can be further improved by using high order thermal Carr-Purcell-Meiboom-Gill (CPMG) measurements \citep{wang2015high,toyli2013fluorescence}, and higher photon collection efficiency \citep{widmann2015coherent,radulaski2017scalable}. A broad-range high sensitivity and nanoscale resolution thermometry can be constructed by using nano-SiC particles, which would be used in a wide variety of system, including living cell and microelectronics systems \citep{kucsko2013nanometre}. Our results are helpful in understanding the spin dynamics in SiC at high temperature and would stimulate further research in this area.

\section*{ACKNOWLEDGMENTS}
We thank Haiou Li for his help in the experiment. This work was supported by the National Key Research and Development Program of China (Grant No. 2016YFA0302700 and 2017YFA0304100), the National Natural Science Foundation of China (Grants No. 61725504, 61327901, 61490711 and 11774335), the Key Research Program of Frontier Sciences, Chinese Academy of Sciences (CAS) (Grant No. QYZDY-SSW-SLH003), Anhui Initiative in Quantum Information Technologies (AHY060300 and AHY020100), the Fundamental Research Funds for the Central Universities (Grant No. WK2470000020 and WK2470000026). This work was partially carried out at the USTC Center for Micro and Nanoscale Research and Fabrication.

F.-F. Y. and J.-F. W. contributed equally to this work.



\begin{thebibliography}{10}

\bibitem{koehl2011room}
W. F. Koehl, B. B. Buckley, F. J. Heremans, G. Calusine, and D. D. Awschalom.
\newblock Room temperature coherent control of defect spin qubits in silicon
  carbide.
\newblock  Nature \textbf{479}, 84 (2011).

\bibitem{falk2013polytype}
A. L. Falk, B. B. Buckley, G. Calusine, W. F. Koehl, V. V. Dobrovitski, A. Politi, C. A. Zorman, P. X-L Feng, and D. D. Awschalom.
\newblock Polytype control of spin qubits in silicon carbide.
\newblock  Nat. Commun. \textbf{4}, 1819 (2013).

\bibitem{christle2015isolated}
D. J. Christle, A. L. Falk, P. Andrich, P. V. Klimov, J. U. Hassan, N. T. Son, E. Janz{\'e}n, T. Ohshima, and D. D. Awschalom.
\newblock Isolated electron spins in silicon carbide with millisecond coherence
  times.
\newblock Nat. Mater. \textbf{14}, 160 (2015).

\bibitem{fuchs2015engineering}
F. Fuchs, B. Stender, M. Trupke, D. Simin, J. Pflaum, V. Dyakonov, and G. Astakhov.
\newblock Engineering near-infrared single-photon emitters with optically
  active spins in ultrapure silicon carbide.
\newblock  Nat. Commun. \textbf{6}, 7578 (2015).

\bibitem{widmann2015coherent}
M. Widmann, S. Y. Lee, T. Rendler, N. T. Son, H. Fedder, S. Paik, L. P. Yang, N. Zhao, S. Yang, I. Booker, et~al.
\newblock Coherent control of single spins in silicon carbide at room
  temperature.
\newblock  Nat. Mater. \textbf{14}, 164 (2015).

\bibitem{castelletto2014silicon}
S. Castelletto, B. Johnson, V. Iv{\'a}dy, N. Stavrias, T. Umeda, A. Gali, and  T. Ohshima.
\newblock A silicon carbide room-temperature single-photon source.
\newblock  Nat. Mater. \textbf{13}, 151 (2014).

\bibitem{lienhard2016bright}
B. Lienhard, T. Schr{\"o}der, S. Mouradian, F. Dolde, T. T. Tran, I. Aharonovich, and D. Englund.
\newblock Bright and photostable single-photon emitter in silicon carbide.
\newblock  Optica \textbf{3}, 768 (2016).

\bibitem{wang2017efficient}
J. F. Wang, Y. Zhou, X. M. Zhang, F. C. Liu, Y. Li, K. Li, Z. Liu, G. Z. Wang, and W. B. Gao.
\newblock Efficient generation of an array of single silicon-vacancy defects in
  silicon carbide.
\newblock Phys. Rev. Appl. \textbf{7}, 064021 (2017).

\bibitem{christle2017isolated}
D. J. Christle, P. V. Klimov, F. Charles, K. Sz{\'a}sz, V. Iv{\'a}dy, Valdas Jokubavicius, Jawad~Ul Hassan, Mikael Syv{\"a}j{\"a}rvi, W. F. Koehl, T. Ohshima, et~al.
\newblock Isolated spin qubits in Sic with a high-fidelity infrared
  spin-to-photon interface.
\newblock Phys. Rev. X \textbf{7}, 021046 (2017).

\bibitem{klimov2015quantum}
P. V. Klimov, A. L. Falk, D. J. Christle, V. V. Dobrovitski, and D. D. Awschalom.
\newblock Quantum entanglement at ambient conditions in a macroscopic
  solid-state spin ensemble.
\newblock Sci. Adv. \textbf{1}, e1501015 (2015).

\bibitem{seo2016quantum}
H. Seo, A. L. Falk, P. V. Klimov, K. C. Miao, G. Galli, and D. D. Awschalom.
\newblock Quantum decoherence dynamics of divacancy spins in silicon carbide.
\newblock Nat. Commun. \textbf{7}, 12935 (2016).


\bibitem{klimov2014electrically}
P. V. Klimov, A. L. Falk, B. B. Buckley, and D. D. Awschalom.
\newblock Electrically driven spin resonance in silicon carbide color centers.
\newblock Phys. Rev. Lett. \textbf{112}, 087601 (2014).

\bibitem{falk2014electrically}
A. L. Falk, P. V. Klimov, B. B. Buckley, V. Iv{\'a}dy, I. A. Abrikosov, G. Calusine, W. F. Koehl, {\'A}. Gali, D. D. Awschalom.
\newblock Electrically and mechanically tunable electron spins in silicon carbide color centers.
\newblock Phys. Rev. Lett. \textbf{112}, 187601 (2014).


\bibitem{zhou2017self}
Y. Zhou, J. F. Wang, X. M. Zhang, K. Li, J. M. Cai, and W. B. Gao.
\newblock Self-protected thermometry with infrared photons and defect spins in silicon carbide.
\newblock Phys. Rev. Appl. \textbf{8}, 044015 (2017).

\bibitem{maze2008nanoscale}
J. R. Maze, P. L. Stanwix, J. S. Hodges, S. Hong, J. M. Taylor, P. Cappellaro, L. Jiang, M. V. Gurudev Dutt, E. Togan, A. S. Zibrov, et~al.
\newblock Nanoscale magnetic sensing with an individual electronic spin in diamond.
\newblock Nature \textbf{455}, 644 (2008).

\bibitem{niethammer2016vector}
M. Niethammer, M. Widmann, S. Y. Lee, P. Stenberg, O. Kordina, T. Ohshima, N. T. Son, E. Janz{\'e}n, and J. Wrachtrup.
\newblock Vector Magnetometry Using Silicon Vacancies in 4 H-SiC Under Ambient Conditions.
\newblock Phys. Rev. Applied \textbf{6}, 034001 (2016).

\bibitem{simin2016all}
D. Simin, V. A. Soltamov, A.V. Poshakinskiy, A. N. Anisimov, R. A. Babunts, D. O. Tolmachev, E. N. Mokhov, M. Trupke, S. A. Tarasenko, A. Sperlich, et al.
\newblock All-optical dc nanotesla magnetometry using silicon vacancy fine structure in isotopically purified silicon carbide.
\newblock Phys. Rev. X \textbf{6}, 031014 (2016).

\bibitem{simin2015high}
D. Simin, F. Fuchs, H. Kraus, A. Sperlich, P. G. Baranov, G. V. Astakhov, and V. Dyakonov.
\newblock High-precision angle-resolved magnetometry with uniaxial quantum centers in silicon carbide.
\newblock Phys. Rev. Applied \textbf{4}, 014009 (2015).


\bibitem{wolfowicz2018electrometry}
G. Wolfowicz, S. J. Whiteley, D. D. Awschalom.
\newblock Electrometry by optical charge conversion of deep defects in 4H-SiC.
\newblock Arxiv. \textbf{1803}, 05956 (2018).


\bibitem{brites2012thermometry}
C. D. Brites, P. P. Lima, N. J. Silva, A. Mill{\'a}n, V.S.
  Amaral, F. Palacio, and L.D. Carlos.
\newblock Thermometry at the nanoscale.
\newblock Nanoscale \textbf{4}, 4799 (2012).

\bibitem{kucsko2013nanometre}
G. Kucsko, P.C.Maurer, N. Y. Yao, M. Kubo, H. J. Noh, P. K. Lo, H.
  Park, and M. D. Lukin.
\newblock Nanometre-scale thermometry in a living cell.
\newblock Nature \textbf{500}, 54 (2013).


\bibitem{toyli2012measurement}
D. M. Toyli, D. J. Christle, A. Alkauskas, B. B. Buckley, C. G. Van. de Walle, and D. D. Awschalom.
\newblock Measurement and control of single nitrogen-vacancy center spins above
  600 K.
\newblock Phys. Rev. X \textbf{2}, 031001 (2012).

\bibitem{plakhotnik2014all}
T. Plakhotnik, M. W. Doherty, J. H. Cole, R. Chapman, and N. B. Manson.
\newblock All-optical thermometry and thermal properties of the optically
  detected spin resonances of the nv--center in nanodiamond.
\newblock Nano Lett. \textbf{14}, 4989 (2014).

\bibitem{li2017temperature}
C. C. Li, M. Gong, X. D. Chen, S. Li, B. W. Zhao, Y. Dong, G. C. Guo, and F. W. Sun.
\newblock Temperature dependent energy gap shifts of single color center in
  diamond based on modified varshni equation.
\newblock Diam. Relat. Mater. \textbf{74}, 119 (2017).

\bibitem{wang2015high}
J. F. Wang, F. P. Feng, J. Zhang, J. H. Chen, Z. C. Zheng, L. P. Guo, W. L. Zhang, X. R. Song, G. P Guo, L. L. Fan, et~al.
\newblock High-sensitivity temperature sensing using an implanted single
  nitrogen-vacancy center array in diamond.
\newblock Phys. Rev. B. \textbf{91}, 155404 (2015).

\bibitem{taylor2008high}
J. M. Taylor, P. Cappellaro, L. Childress, L. Jiang, D. Budker, P. R. Hemmer, A. Yacoby, R. Walsworth, and M. D. Lukin.
\newblock High-sensitivity diamond magnetometer with nanoscale resolution.
\newblock Nat. Phys. \textbf{4}, 810 (2008).

\bibitem{neumann2013high}
P. Neumann, I. Jakobi, F. Dolde, C. Burk, R. Reuter, G. Waldherr, J. Honert, T. Wolf, A. Brunner, J. H. Shim, et~al.
\newblock High-precision nanoscale temperature sensing using single defects in
  diamond.
\newblock Nano Lett. \textbf{13}, 2738 (2013).

\bibitem{schirhagl2014nitrogen}
R. Schirhagl, K. Chang, M. Loretz, and C. L. Degen.
\newblock Nitrogen-vacancy centers in diamond: nanoscale sensors for physics
  and biology.
\newblock Annu. Rev. Phys. Chem. \textbf{65}, 83 (2014).

\bibitem{toyli2013fluorescence}
D. M. Toyli, F. Charles, D. J. Christle, V. V. Dobrovitski, and D. D. Awschalom.
\newblock Fluorescence thermometry enhanced by the quantum coherence of single
  spins in diamond.
\newblock Proc. Natl. Acad. Sci. USA \textbf{110}, 8417 (2013).

\bibitem{anisimov2016optical}
A. N. Anisimov, D. Simin, V. A. Soltamov, S. P. Lebedev, P. G. Baranov, G. V. Astakhov  and V. Dyakonov.
\newblock Optical thermometry based on level anticrossing in silicon carbide.
\newblock  Sci. Rep. \textbf{6}, 33301 (2016).


\bibitem{radulaski2017scalable}
M. Radulaski, M. Widmann, M. Niethammer, J.L. Zhang, S. Y. Lee, T. Rendler, K. G. Lagoudakis, N. T. Son, E. Janzen, T. Ohshima, et~al.
\newblock Scalable quantum photonics with single color centers in silicon
  carbide.
\newblock Nano Lett. \textbf{17}, 1782 (2017).

\end{thebibliography}
\end{document}